\documentstyle[aps,psfig,times]{revtex}
\begin{document}
\title{Robust gamma oscillations  in networks of inhibitory Hippocampal 
interneurons.}
\author{P.H.E. Tiesinga$^{\sharp,\diamondsuit,}$\footnote{Corresponding 
author,\\Tel: 619 453 4100 ext 1039\\ Fax: 619 455 7933\\ E-mail:
 tiesinga@salk.edu}  
and Jorge V. Jos\'e $^{\sharp}$\\
$^\sharp$ Center for Interdisciplinary Research on Complex Systems,
and Department of Physics, Northeastern University, 360 Huntington Ave,
Boston, Massachusetts 02115, USA\\
$^\diamondsuit$ Sloan Center for Theoretical Neurobiology, Salk Institute,\\
 10010 N. Torrey Pines Rd.,
La Jolla, CA 92037. }
\maketitle
\begin{abstract}
Recent experiments suggest that inhibitory networks of interneurons
can synchronize the neuronal discharge in {\em in vitro} hippocampal slices.
Subsequent theoretical work has shown that strong synchronization by 
mutual inhibition is only moderately robust against neuronal heterogeneities 
in the current drive, provided by activation of metabotropic glutamate 
receptors.  {\it In vivo} neurons display greater variability in the 
interspike intervals  due to the presence of synaptic noise.
Noise and heterogeneity affect synchronization properties differently.
In this paper we study using model simulations 
how robust synchronization can  be 
in the presence of synaptic noise and neuronal heterogeneity.
We find that with at least a  minimum amount of noise 
stochastic weak synchronization (SWS) (i.e. when neurons spike within a short
interval from each other, but not necessarily at each period) is produced that
is much more robust than strong synchronization 
(i.e. when neurons spike each period).  
The statistics of the SWS population discharge are consistent 
with previous experimental data.
We find robust SWS in the gamma frequency range ($20$-$80$ Hz) 
for a stronger synaptic coupling compared to previous models and
for networks with $10$ -- $1000$ neurons. 

\vspace*{0.5in}

\end{abstract}
\newpage
\twocolumn
\section{Introduction}
One of the important properties of the behavior of the nervous system,
 discovered early on  \cite{Adrian50}, that has attracted
significant amount of attention, is the synchronization 
of neuronal discharges.  In recent years the advent of improved 
experimental techniques has provided  vast amounts of new synchronization
data. Concomitantly, there has been a resurgence in interest and 
controversy concerning the functional relevance of synchronization. 
It has been established that {\it in vivo} cortical  neurons have noisy 
spike trains \cite{Softky93} (but see in contrast \cite{Gur97}),
and that groups of neurons discharge coherently as found  in population 
recordings (such as EEGs, or by arrays of extracellular electrodes, for 
a review see \cite{Aertsen93}). These two facts have sparked major 
controversies. Firstly, does noise (or precise timing) in neuronal spike 
trains contain  information \cite{Rieke97,Softky95}, or is information merely 
due to noisy processing of an average firing rate 
\cite{Shadlen94,Shadlen95,Shadlen98}? Secondly, is synchronization
functionally (or even statistically) significant\cite{Laurent97,Abeles98}, 
or just an epiphenomenon \cite{Shadlen98}? 
In this paper we focus on two different aspects of synchronization
that have received little attention so far.  Can realistic neuronal networks 
synchronize under the biological conditions of variable intrinsic 
neuronal properties, and the noise-induced neuronal unreliability?
What kind of synchronization can be obtained, and what are 
its pertinent statistical properties? It is necessary
to resolve these  two questions  to properly  formulate the issues 
to be studied in experiment, and to analyze different
 ways of probing the experimental 
data. Here we focus our attention on the extensively-studied synchronous 
gamma oscillations in hippocampus  
\cite{traub,Traub98a,whittington,traub2,buhl98,JM99}.
Theoretical and computational work has shown that mutual inhibition
is capable of synchronizing neuronal networks \cite{vreeswijk94,Wang93}. 
Subsequent  in vitro experiments have convincingly established
the role of GABA-ergic hippocampal interneurons in gamma oscillations 
\cite{whittington,traub}. Wang \& Buzs\'aki \cite{wang} studied the effect of 
current heterogeneity and partial connectivity on the synchronization 
of the hippocampal network. They only found strong synchronization 
in the gamma frequency range when the current heterogeneities 
were small \cite{wang,white}. In strong synchronization all neurons
in a local circuit spike within a short interval of each other.
This suggests that strong synchronization can only be obtained when 
the intrinsic properties of the neurons are not too different.
According to Ref.\cite{wang} this would mean a less than $10\%$ difference in 
current drive, or average firing rate. It has been hard to pinpoint the 
amount of variability  in intrinsic properties in the {\em in vitro}
and {\it in vivo} preparations of different brain areas.
It is however not unreasonable to assume the presence of more than 
$10\%$ variability in these preparations. Strong
synchronization is also not robust against noise \cite{CNS}.
It would therefore seem unlikely for strong synchronization
to be present in hippocampus under physiological conditions.
Indeed, here we show that stochastic weak synchronization (SWS)
is more prevalent in parameter space, and is also robust against
neuronal heterogeneities and synaptic noise.  We conjecture that 
as a consequence  it is much more likely to occur in neuronal systems. 

In SWS, neurons spike within a short interval from each other, 
but not necessarily at each period \cite{Hakim98,Amit97}. 
The synchronization is called stochastic, because
the particular cycle in which the neurons fire is random.
This makes the properties of this state different
from the well known cluster states studied by previous authors
\cite{Golomb92,GolombRinzel93,GolombRinzel94,Kopell94}.
There each neuron always fires at the same cycle with the same
cluster.  Both strong and stochastic weak synchronization 
yield periodic population oscillations. The difference  
can then only be ascertained using multi-unit recordings. 

We use cross correlation analysis to show that noise and heterogeneity 
affect the  synchronization properties of our network in very different 
ways. Large enough noise and heterogeneity will, however,  stop 
strongly synchronized oscillations. We demonstrate that neither 
adding a periodic drive nor increasing synaptic coupling can significantly 
increase robustness of strong synchronization. Finally we determine 
for what parameters robust self-induced 40Hz synchronous oscillations 
can be obtained.

\section{Methods}
\subsection{Single neuron model}
Our aim here is to establish physiological
criteria for robust synchronization in the gamma frequency range.
The use of a biophysically realistic model is therefore of pivotal importance.
At the same time it is also important to balance the amount of complexity
versus practical simplicity \cite{Rall95}.
We have therefore not attempted to use the latest available data to construct
a detailed multi-compartmental model. The computer requirements
to sample the full relevant parameter space, and perform our type of
analysis,  would be extremely demanding even using very fast 
computers.  
It has been shown, nonetheless,  that one and two compartmental models can
accurately generate spike trains of the right shape and frequency
\cite{PinskyRinzel94,Mainen94,MainenSejn98}.
Multi-compartmental models may be necessary to assess the synaptic 
integration of inputs located on different parts of the dendritic
tree. This is currently an intensely studied area in
electrophysiology \cite{Yuste96,Koch97,Sakmann99}. 
Here we study a model  previously introduced by others\cite{wang}. The model 
has been shown to reproduce the salient features of the dynamics
of hippocampal interneurons.
The neurons are modeled as a single compartment
with Hodgkin-Huxley type sodium and potassium channels.
In this work all the neurons are connected to all others and
themselves (ALL to ALL connectivity) via
inhibitory GABA$_A$-synapses. 
The equation for the membrane potential of a neuron is 
(the index $i$ of the neuron is omitted)
\begin{equation}
C_m \frac{dV}{dt}=-I_{Na}-I_K-I_L-I_{syn}+I+C_m\xi.
\label{SINGNEUR}
\end{equation}
Here we use: the leak current
\(
I_L=g_L (V-E_L),
\)
the sodium current
\(
I_{Na}=g_{Na} m_{\infty}^3 h (V-E_{Na}),
\)
the potassium current:
\(
I_K=g_K n^4 (V-E_{K}),
\) and the synaptic current:
\(
I_{syn}=g_{syn}s(V-E_{syn})
\).
The Gaussian noise is denoted as $\xi$ (see below), 
and $I$ is the tonic drive.
The channel kinetics are given in terms of $m$, $n$, and $h$. 
They satisfy the following first order kinetics:
\begin{equation}
\frac{dx}{dt}=\phi (\alpha_x (1-x)-\beta_x x).
\end{equation}
Here $x$ labels the different kinetic variables $m$, $n$, and $h$, and 
$\phi=5$ is a dimensionless time-scale that 
can be used to tune the temperature-dependent 
speed with which the  channels open
or close. The rate constants 
are \cite{wang},
\begin{eqnarray}
\alpha_m&=& \frac{-0.1(V+35)}{\exp(-0.1(V+35))-1}, \nonumber\\
\beta_m&=& 4\exp(-(V+60)/18), \nonumber\\
\alpha_h&=&0.07\exp(-(V+58)/20), \nonumber\\
\beta_h&=&\frac{1}{\exp(-0.1(V+28))+1}, \nonumber\\
\alpha_n&=&\frac{-0.01 (V+34)}{\exp(-0.1(V+34))-1}, \nonumber\\ 
\beta_n&=&0.125~\exp(-(V+44)/80). \nonumber
\end{eqnarray}
We make the approximation that $m$ follows the asymptotic value
$m_{\infty}(V(t)))=\alpha_m/(\alpha_m+\beta_m)$ instantaneously.
The synaptic gating variable $s$ obeys the following equation 
\cite{Perkel81,wangrinzel93,wang}:
\begin{equation}
\frac{ds}{dt}=\alpha F(V_{p})(1-s)-\beta s,
\label{SYNNEUR}
\end{equation}
with $\alpha=12~ ms^{-1}$, $\beta=1/\tau_{syn}$, 
$F(V_p)=1/(\exp(-V_p/2)+1)$, and $V_p$ is the presynaptic potential.
The function $F(V_p)$ is chosen such that when the presynaptic
neuron fires, $V_p>0$, the synaptic channel opens.
The decay time of the postsynaptic hyperpolarization
is chosen as $\tau_{syn}=1/\beta= 10~ms$ (or $20$ ms in some instances).
We use a reversal potential of 
$E_{syn}=-75~mV$ for the inhibitory (GABA$_A$) synapses \cite{Buhl95}. 
The standard set of values for the conductances used in this work is 
$g_{Na}=35$, $g_K=9$, $g_L=0.1$, and $g_{syn}=0.1$ (in $mS/cm^2$),
and we have taken 
$E_{Na}=55~mV$, $E_K=-90~ mV$, and $E_L=-65~ mV$. The membrane
capacitance is $C_m=1\mu F/cm^2$. 
Unless stated otherwise we will use the standard set
of parameters listed above. When no current
value is specified we use $I=1\,\mu A/cm^2$. The network will then
spike at approximately $39$ Hz.\\
We chose the initial values for the membrane potential at the start of 
the simulations uniformly random between $-70$ and $-50~mV$.
The kinetic variables $m$, $n$, $h$, and $s$ are set 
to their asymptotic stationary values 
corresponding to that starting  value of the membrane potential.

The resulting equations with noise are integrated using an adapted
second order Runge-Kutta method \cite{greenside}, with time step $dt=$0.01 ms. 
The accuracy of this integration method was checked for 
the dynamical equations without noise ($D=0$) by
varying $dt$ and comparing the result to 
the one obtained with the  standard 4th order Runge-Kutta method \cite{Press} 
with a time-step $dt$ of $0.05~ms$. 

We normalize all quantities by the surface area of the neuron.
This leads to the following system of units: the membrane
potential $V$ in $mV$, time $t$ in ms, firing rate $f$ in Hz,
membrane capacitance $C_m$ in
$\mu F/cm^2$, conductance $g_x$ in $mS/cm^2$, voltage noise $\xi$ in $mV/ms$,
strength of neuroelectric noise $D$ in $mV^2/ms$,
the rate constants $\alpha_x$ and $\beta_x$ in $ms^{-1}$, and the current $I$
in $\mu A/cm^2$. The kinetic variables $m$, $n$, $h$, $s$, and the 
time-scale $\phi$ are dimensionless. Results in our paper are expressed in
this system of units. 

\subsection{Heterogeneity and synaptic noise}

We have included heterogeneity in the applied current.
For each run we draw the applied current for each neuron 
from a uniform distribution. The average of the current distribution 
is $I$ and the variance is $\sigma^2_I$.
The current heterogeneity represents the variation
in the intrinsic properties of the neurons in the hippocampus.
Experimental measurements of quantities like the input resistance $R_{in}$,
the membrane time-scale, the spontaneous spiking rate,
the shape of the somatic action potential (amplitude, width, rise and 
fall time), and the afterhyperpolarization, show considerable variance 
\cite{Lacaille87,Lacaille90,Morin96}.  
It is hard to determine how much of the variance is due to measurement
errors, and how much is actually due to intrinsic neuronal variability.
Here we assume the main effect of the variability is to
change the intrinsic frequency of the neurons (which can be varied
using the current drive in our model).
Another source of heterogeneity in {\it in vitro} experiments is 
the glutamate pressure
ejection method \cite{whittington}. It can lead to an inhomogeneous 
activation of metabotropic glutamate receptors, and thus to a variable current.
In this paper we will consider $\sigma_I$ as a free parameter.

At least three sources of noise can be identified \cite{JohnstonWubook}:  
random inhibitory postsynaptic potentials (IPSP) and excitatory postsynaptic
potentials (EPSP), stochasticity of the synaptic transmission,
and the stochasticity of the channel dynamics. 
Here we assume that the variability in the neuronal discharge is
mainly due  to synaptic noise \cite{Calvin67}.
We have compared the effects 
of Poisson distributed spike trains of EPSPs and IPSPs to that of a Gaussian
noise current on interspike interval (ISI) variability. Poisson and Gaussian noises, 
do not yield identical results.
The statistics obtained from both models, however, are similar in the parameter
regime studied \cite{CNS98}. For
the purpose of our studies we  consider that Poisson and Gaussian
distributions
are two alternate ways of producing noisy spike trains with particular
statistics. Therefore, the synaptic  
noise is only implemented as a Gaussian distributed,
white noise current in neuron $i$, with $\langle \xi_i(t) \rangle=0$, and 
$\langle \xi_i(t) \xi_j(t') \rangle = 2D \delta(t-t')\delta_{ij}$. 
The noise currents in different neurons are assumed independent.

\subsection{Measured quantities} 
From our simulations we obtain the time trace for the
membrane potential $V_i(t)$ of each neuron. We determine
the spike-trace $X_i$ from $V_i$ as follows: $X_i(t)=1$ when $V_i(t)$
 crosses $0\,mV$ (i.e. $V(t^-)\le 0<V(t^+)$), and it is zero elsewhere. 
From $X_i$ we obtain $X(t)=\sum_i X_i(t)$. $X$ is proportional 
to the instantaneous firing rate of the network. We also calculate 
the correlations function $\kappa$ \cite{wang}:
\begin{equation}
\kappa\equiv \sum_{i\ne j} \frac{\langle 
{\hat X}_i(n\tau){\hat X}_j(n\tau)\rangle }
{\sqrt{\langle {\hat X}_i(n\tau)\rangle  \langle {\hat X}_j(n\tau)\rangle}}.
\label{calefaccion}
\end{equation}
This function measures the amount of strong synchronization, 
and depends on the bin size $\tau$ of the time discretization
\begin{equation}
{\hat X}_i(n\tau)=\int_{(n-1)\tau}^{n\tau} ds~ X_i(s).
\end{equation}
We use $\tau=200~dt=2~ms$ for oscillations in the gamma-frequency 
range,  or $T/10$ for periodic drives with period $T$.

We also evaluate other measures that yield further
detailed quantitative characterization of the network behavior.
We calculate the time autocorrelation function:
\begin{equation}
g_x(t)= \frac{\langle x(t) x(0) \rangle-\langle x \rangle^2}{
(\langle x^2\rangle-\langle x \rangle^2)},
\end{equation}
and the cross correlation function:
\begin{equation}
g_{xy}(t)= \frac{\langle x(t) y(0) \rangle-\langle x(t) 
\rangle \langle y(0) \rangle}
{\sqrt{\langle x(t)^2\rangle \langle y(0)^2\rangle}}.
\end{equation}
Here $x$ and $y$ can be any of the variables
$X_i$, $V_i$, and $X$,
and $\langle \rangle$ is a shorthand notation for the time-average.

We also consider the more conventional interspike interval 
histogram (ISIH) \cite{Gerstein62}, averaged over all network neurons.
From the ISIH one can obtain two statistics: the average ISI,
$\tau_{ISI}$, and the standard deviation of the ISI, $\sigma_{ISI}$. The
ratio $\sigma_{ISI}/\tau_{ISI}$ is known as the coefficient of variation (CV).
The average firing rate is $f=1/\tau_{ISI}$, and the population standard
deviation of $f$ is $\sigma_f$. 
\begin{equation}
\sigma_f=\sum_j f_j^2 - f^2,
\end{equation}
where $f_j=1/\tau_{ISI}^j$ is the average firing rate of the $j$th neuron.
In addition we plot  rastergrams, with the action potential of each neuron
plotted as a filled circle, with the y-coordinate given by the neuron index 
and the x-coordinate by the spiking time.

To analyze the stochastic weak synchronization network 
dynamics we need to apply a different method. 
The population period $\tau_n$ is different from the population 
averaged ISI, and to estimate it we proceed as follows.
First we determine the firing rate ${\hat X}(t)$ as before 
with $1$ ms bins. In the stochastic weak synchronization 
state ${\hat X}(t)$ will consist of a 
number of approximately equidistant peaks of finite width
(see Fig.$~$\ref{EXTRA5}f). We use the position of the first 
maximum of the Fourier transform at nonzero frequency as an estimate $T$ for the period  
$\tau_n$. We calculated the weight $\langle {\hat X}(t) \rangle$,
the average position $t_c^i=\langle t{\hat X}(t)\rangle$,
and the width $\sigma_c^i=\sqrt{\langle t^2 {\hat X}(t)\rangle-(t_c^i)^2}$ 
of the $i$th peak. The time average is taken over a range 
$[-0.35 T, 0.35 T]$ about the estimated position
$t_c^{i-1}+T$ of the peak. We calculated the number of spikes that fall outside
this region. If the average number of missed spikes is more than one 
per cycle we reject the cluster state. The cycle length (time between 
two consecutive cluster firings) is  defined as $\tau_c^i=t_c^i-t_c^{i-1}$.
We determine the average cluster size $N_c=\langle N_c^i\rangle$, its
$CV(N_c)=\sqrt{\langle (N_c^i)^2\rangle-N_c^2}/N_c$, the average cycle
length $\tau_n=\langle \tau_c^i \rangle$,  
its $CV(\tau_n)=\sqrt{\langle (\tau_c^i)^2\rangle-\tau_n^2}/\tau_n$,
and the average width $\sigma_c=\langle \sigma_c^i\rangle$. Here the 
average $\langle \cdot \rangle$ is given by the sum 
over all cycles in the run (after discarding a transient).
We characterize the strength of the synchronization using a 
modified  $\kappa_W$ and $CV_W$. In the SWS state the ISIH has multiple 
peaks. The CV of the ISI receives contributions
from the variance within each peak, but also of the variance between 
the multiple peaks. We are only interested in the former, and the 
conventional CV is thus an overestimate. Instead we use 
$CV_W=\sigma_c/\tau_n$ which is related to the average width of one peak
in the ISIH. The coherence $\kappa$ measures the number of 
coincident spikes between  two spike trains. Consider two neurons that do 
not spike at each cycle, but when they
both do, the spikes are coincident (that is in 
the same bin).
If the probability of spiking in a cycle is $p=N_c/N$, and both neurons
fire statistically independent, we obtain $\kappa=p$. These neurons 
can be considered synchronous
and we want $\kappa_{W}=1$. We therefore normalize $\kappa$ by $p$. 

There is a subtlety in the calculation of the average firing rate. In the
deterministic noiseless case one ISI is enough to determine the average value 
(after discarding the transient).
[Note that counting the number of spikes  in a fixed interval is not an 
efficient way to determine the exact firing rate.] In the 
presence of noise, however, you need at least $10$
ISIs to accurately determine the average. In networks with large 
current heterogeneities  there are neurons with high and very low firing 
rates (Fig.$~$\ref{EXTRA5}). The average ISI for the low firing rate is 
less accurate than for the high firing rate neurons in the network.
However, it carries equal weight in the  conventional average $\tau_{ISI}=\sum_j 
\tau_{ISI}^j$. We have therefore used a weighted average 
$\tau_{ISI}=\sum_j n_j \tau_{ISI}^j/\sum_j n_j$
(here $n_j$ is the number of intervals over which 
$\tau_{ISI}^j$ is calculated), and the approximate 
identity $N_c/\tau_n\approx N_s/\tau_{ISI}$ can be used as a check.
$N_s$ is the number of active neurons, defined as the neurons that 
have more than two ISIs  after the transient.

\section{Results}
\subsection{Non robustness of strong synchronization}
\begin{figure*}
\newpage
\unitlength=0.1in
\begin{picture}(20,50)
\includegraphics{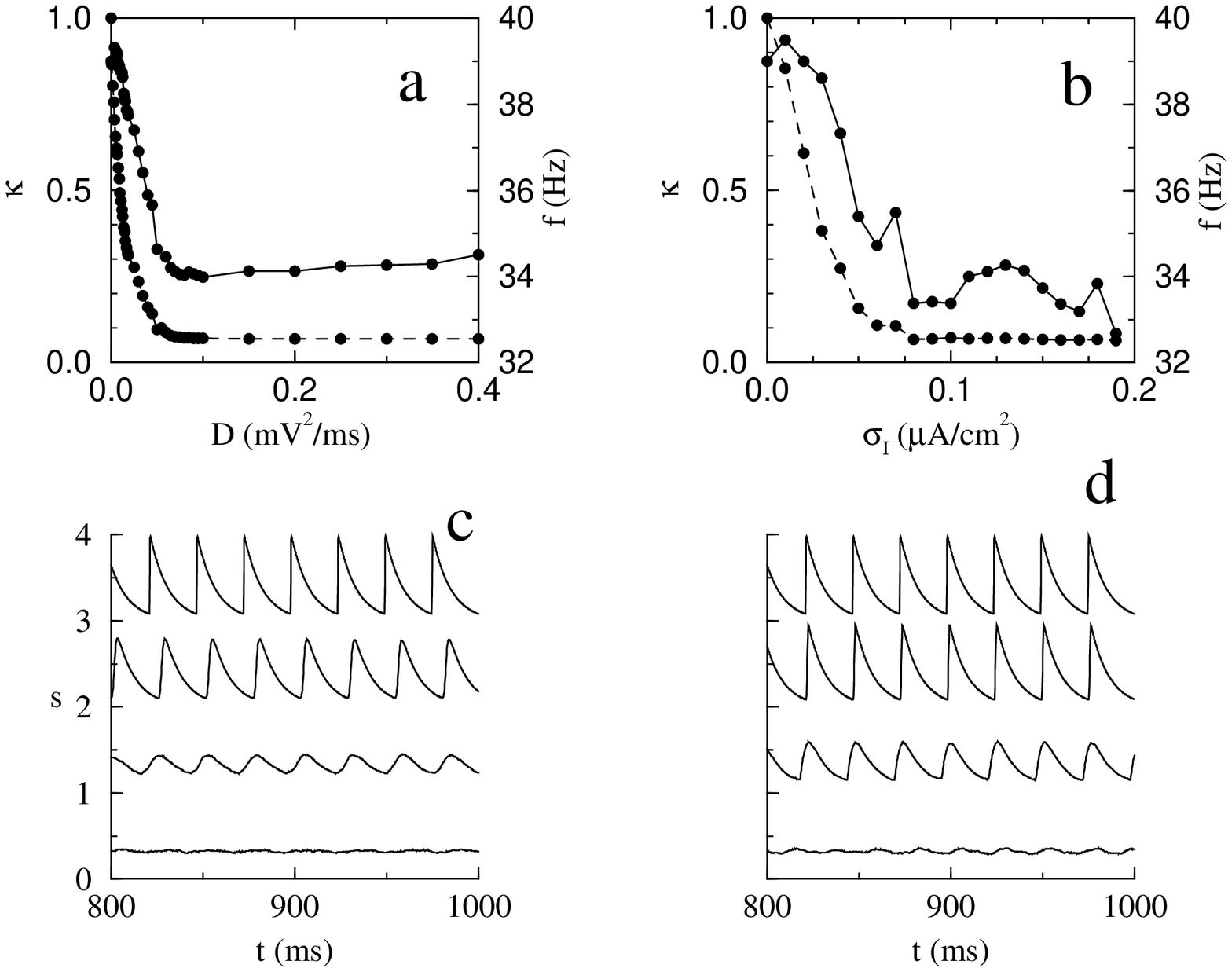}
\end{picture}
\caption{Coherence parameter $\kappa$ (dashed line, left hand scale),
and average network frequency $f$ (continuous line, right hand scale)
versus (a) noise strength $D$ (with $\sigma_I=0$)
and (b) current heterogeneity $\sigma_I$ (with $D=0$). 
After a transient of $1~s$, (a)
time-averages were computed over $3~s$, and (b) $2~s$. In (b)
each point represents the results of one independent value
of the current heterogeneity. In (c,d) we plot the synaptic drive $s(t)$,
(c) for $\sigma_I=0$ with $D=0$, $0.01$, $0.04$, and $0.09$ from top to bottom;
whereas in (d) $D=0$ with $\sigma_I=0$, $0.01$, $0.04$, 
and $0.09$ from top to bottom. The  curves are  offset by multiples of $\Delta s=1$.
 We used the standard set of parameters described in Methods.
}
\label{RFIG5}
\end{figure*}

\begin{figure*}
\newpage
\unitlength=0.1in
\begin{picture}(40,70)
{
\includegraphics{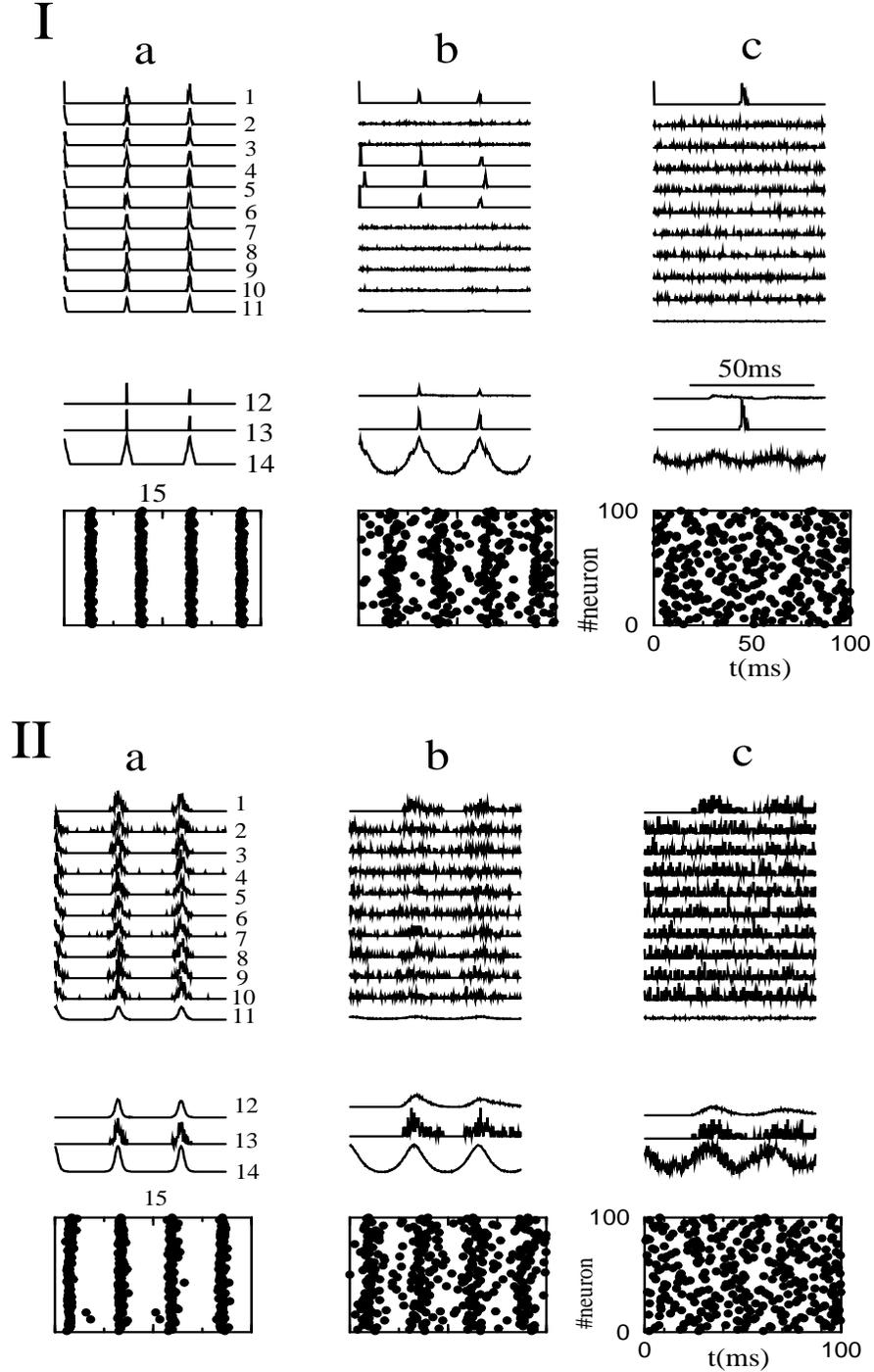}
}
\end{picture}
\vspace*{0.8in}
\caption{(I) Loss of synchrony due to current heterogeneities:
(a) $\sigma_I=0.02$, (b) $0.07$, (c) $0.08$, respectively.
(II) Loss of synchrony due to synaptic noise:
$D=0.01$ (a), $0.04$ (b), $0.09$ (c), respectively.
In $1$ and $13$ we plot  the autocorrelation of $X_1$ 
(spike train of neuron $1$); In
$2-10$ we plot the cross correlation of $X_1$ with $X_2,\cdot\cdot ,~X_{10}$, 
respectively; In $11$ we show the cross correlations between 
$X_i$ and $X_j$, averaged over all pairs $i,j$; In
$12$ we plot the autocorrelation of $X_i$, averaged over all neurons, and in
$14$ the auto correlation of the total spiking rate $X$;
In $15$ we show the rastergrams of the network, i.e. neuron number 
versus spiking time. The time-scale bar, shown in Ic12, applies to 
the curves $1-14$, (a-c) in I and II.
The y-axis is in arbitrary units, and the same scale is used for the 
curves in $1-11$, and $12-14$, except for the curves IIa 12,13 and IIb 12,13,
which are rescaled by a factor of $10$. 
After a transient of $1~s$, the time averages are computed over 
$2~s$ (I), and $3~s$ (II).
 We used the standard set of parameters described in Methods.
}
\label{RFIG6}
\end{figure*}

\begin{figure*}
\newpage
\unitlength=0.1in
\begin{picture}(20,50)
\includegraphics{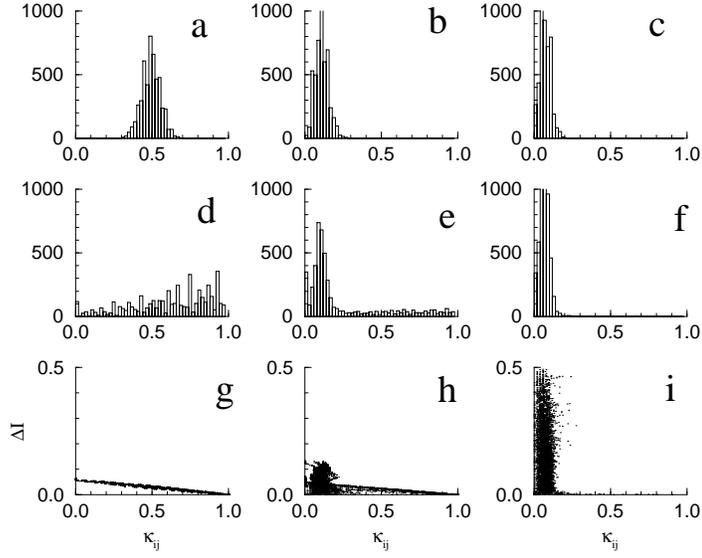}
\end{picture}
\caption{(a-f) distribution of $\kappa$ values for individual pairs ($i$,$j$),
(g-i) difference in driving current $\Delta I=\vert I_i-I_j\vert$ versus
$\kappa$. In (a-c) $\sigma_I=0$ and (a) $D=0.01$, (b) $0.04$, 
(c) $0.09$.
Whereas in (d-i) $D=0$ and (d,g) $\sigma_I=0.02$, (e,h) $0.04$, and 
(f,i) $0.15$.
A transient of $500$ ms was discarded before averaging over $2000$ ms.
 We used the standard set of parameters described in Methods.
}
\label{EXTRA1}
\end{figure*}

\begin{figure*}
\newpage
\unitlength=0.1in
\begin{picture}(40,50)
{
\includegraphics{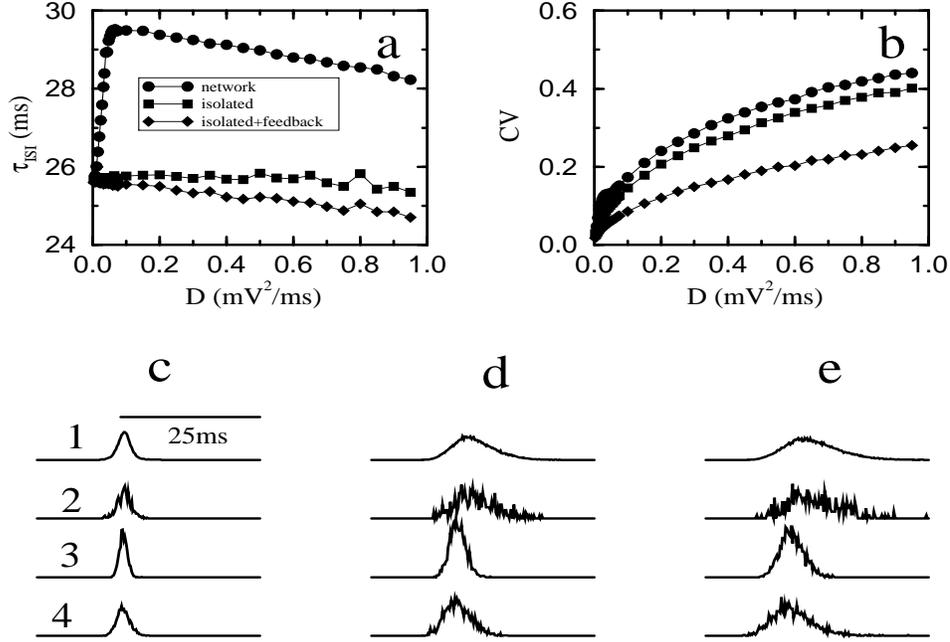}
}
\end{picture}
\caption{(a) Average interspike interval $\tau_{ISI}$,
and (b) coefficient of variation (CV), plotted
versus noise strength $D$.  We consider four different cases of ISIHs, 
labeled by numbers in (c)-(e) and by symbols in (a) and (b),
(1) (circles) averages over all neurons in the network,
(2) one neuron in the network,
(3) (diamonds) isolated neuron with autosynaptic feedback,
and (4) (squares) isolated neuron without feedback. 
The noise strengths are (c) $D=0.01$, (d) $0.04$, and (e) $0.09$,
 respectively. In all graphs $I=1.0$ except in case (4), with $I=0.61$.
The time-scale bar for the ISIH is shown in c 1, the y-scale is arbitrary, 
but it is the same for all curves with the same $D$-value.
The averaging times are $10~s$ (network), and $200~s$ (isolated neuron).
}
\label{RFIG7}
\end{figure*}

In this section we describe the results of our simulations
for a network of $N=100$ interneurons, connected all to all, with either
synaptic noise (SN), or current heterogeneities (CH).
In Fig.$~$\ref{RFIG5} we plot coherence parameter $\kappa$ (defined in Eq.$~$(\ref{calefaccion}))
versus the strength of the synaptic noise D, and versus the standard deviation
of the current heterogeneities $\sigma_I$. We find that strong
synchronization is lost for approximately $D>0.10~mV^2/ms$ and 
$\sigma_I > 0.1~\mu A/cm^2$  (with the
standard set of parameters listed in Methods). 
The mechanism by which strong synchronization is lost, however, is different
in the CH case compared to the mechanism with SN. This difference shows up only
if we studies the whole state of the network using cross 
correlation functions, instead of the average quantities shown 
in Fig.$~$\ref{RFIG5}. Next we compare these 
two mechanisms. Wang\&Buzsaki (WB) \cite{wang} have already analyzed 
the case with current heterogeneity. We have reproduced part of their 
work, and we  will refer to their corresponding figures.
In both CH and SN cases the neuronal firing rate decreases when the network 
desynchronizes.  We have plotted the time-trace of the synaptic 
drive $s(t)$ in Figs.$~$\ref{RFIG5}c and d. The phasic part decreases, 
and the tonic part of $s(t)$ increases with increasing $D$ and $\sigma_I$.
The increased tonic part is responsible for the lower average firing rate.
The firing rate of the CH neurons saturates 
(when averaged over enough realizations of the current
heterogeneities\footnote{This assumes 
\[ \int_{I_{av}-\sigma_I\sqrt{12}}^{I_{av}+\sigma_I\sqrt{12}} 
f(I) dI\approx f(I_{av})\]}),  whereas for the SN it increases 
steadily as a function of $D$ for large values of $D$.  This is because 
the single 
neuron firing rate increases with $D$  \cite{CNS98}, but tonic
inhibition saturates to its highest value in the asynchronous network. 
The dispersion $\sigma_f$ (see Methods)
with CH is larger than the  one in SN (not shown).
In SN all the neurons have identical intrinsic properties, and
the expectation value for the average frequency of each neuron 
is the same. The dispersion $\sigma_f$ in this case represents the 
fluctuations in the average ISI due to the finite averaging time. With CH
the neurons have different intrinsic frequencies,
and the dispersion $\sigma_f$ increases with $\sigma_I$ (and does not go 
to zero after a long averaging time, see WB Fig.$~$5B).
In Fig$~$\ref{RFIG6}I and II we compare the correlation functions for the
SN and CH case, respectively. In (a) we have the strongly 
synchronous network, in (c) the asynchronous network, and in (b) a 
transition state. The difference between SN and CH becomes clear 
when one considers the cross correlation functions. With CH the number of pairs
that are phase locked drops gradually (see WB Fig.$~$8E). The pairs that 
are phase locked, are tightly phase locked (Fig.$~$\ref{RFIG6}Ib1,4,5,6), and
there is no dispersion in the cross correlations,
only a relative phase. Even in the asynchronous state the autocorrelation
function $g_{X_i}$ for a single neuron is sharp, i.e. the neuron fires
regularly with a fixed frequency (Fig.$~$\ref{RFIG6}Ic1 and c13).
The population average of $g_{X_i}$, however,
is disordered (Fig.$~$\ref{RFIG6}Ic12), since each of the neurons has a different firing rate.
In the SN case there is already dispersion due to the noise-induced jitter in the spike time, in the autocorrelation Fig.$~$\ref{RFIG6}IIa11, 
and in the cross correlations (Fig.$~$\ref{RFIG6}IIa2-a10). 
The dispersion increases gradually with $D$. 
The difference between the CH and SN cases is also evident in the distribution 
of $\kappa$ values for each pair in the network (Fig.$~$\ref{EXTRA1}, 
WB Fig.$~$8E). For SN there is a well defined peak, with the average 
shifting to lower values as $D$
increases, (Fig.$~$\ref{EXTRA1}a-c), whereas for CH there is a broad 
distribution for small $\sigma_I$
(Fig.$~$\ref{EXTRA1}d), a peak
at low values of $\kappa$  combined with a broad distribution for 
moderate values
of $\sigma_I$ (Fig.$~$\ref{EXTRA1}e). For higher values of $\sigma_I$ 
the network is in an asynchronous regime,
and only the peak for low $\kappa$ values is present (Fig.$~$\ref{EXTRA1}f). 
We have compared the ISIH for a network neuron to the ISIH of 
an isolated neuron (Figs.$~$\ref{RFIG7}c-e), and also 
the values of
$\tau_{ISI}$ and $\sigma_{ISI}$ (Figs.$~$\ref{RFIG7}a and b).
The CV of the network neuron is higher 
than the CV of an isolated neuron which in turn is higher than
the CV of an isolated neuron with autosynaptic feedback. The inhibitory
coupling in the network increases the effect of the noise 
compared to uncoupled neurons: the jitter in the spike times reduces
the phasic component of $s(t)$ (Fig.$~$\ref{RFIG5}c). This effect does not 
take place in a neuron with autosynaptic feedback: the size of the 
phasic component
does not decrease with $D$, only the timing deteriorates.\\

\subsection{Weak enhancement of robustness due to resonance effect} 
\begin{figure*}
\newpage
\unitlength=0.1in
\begin{picture}(20,50)
{
\includegraphics{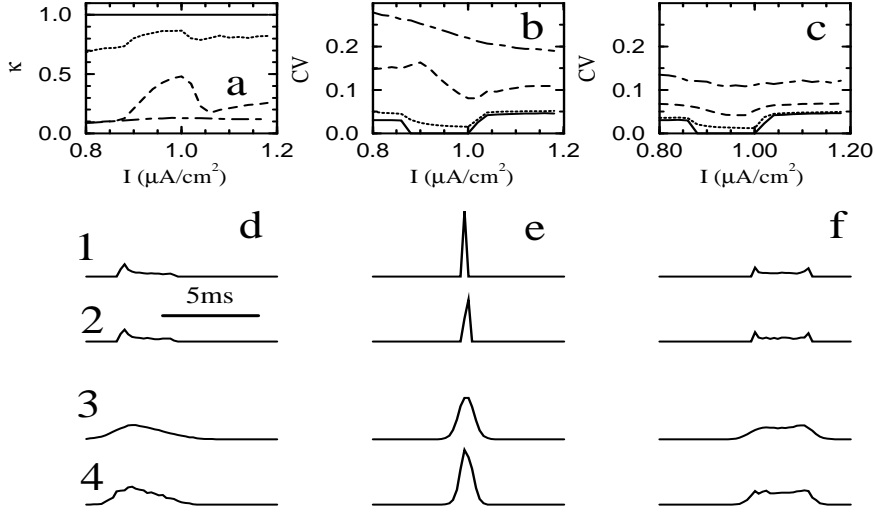}
}
\end{picture}
\caption{(a) Coherence parameter $\kappa$, 
(b) the coefficient of variation CV for a network, and (c) CV for an 
isolated neuron,  versus the applied current $I$ for different values of 
$D=0.0$ (continuous line), $0.004$ (dotted line), $0.04$ (dashed line), and 
$0.2$ (dot-dashed line). ISIHs for a network of neurons (1,3), and  
an isolated  neuron (2,4) for different values of the current 
(d) $I=0.8$, (e) $0.96$, and (f) $1.14$. We used $D=0.0$ (1,2) and 
$0.004$ (3,4). The bar time-scale for  the ISIH is shown in d2. 
We applied an excitatory drive with reversal potential 
$E=0~mV$, strength  $g_{syn}=0.05$, and frequency $40~Hz$.
The isolated neuron has an  autosynaptic feedback of $g_{syn}=0.1$. 
After a transient of $1~s$ ($0.5~s$), time-averages were calculated 
over $10~s$ ($50~s$) for the network (isolated neuron).
}
\label{RFIG8}
\end{figure*}

In this section we drive the tonically active neurons 
with an external periodic drive.
This drive may represent the effects of putative pacemaker
neurons, similar to the ones that were recently found in the striate cortex 
of cats \cite{Gray96}. There is no compelling evidence for having the $40$Hz pacemaker 
neurons projecting to hippocampal interneurons.
Here we model the pacemaker as an
excitatory synapse driven by a periodic pulse train, and investigate
its effects on the robustness of synchronization.
When a single neuron is driven by a periodic drive
it will entrain, or phase lock, when the drive frequency
is close to the natural frequency (or in some cases close to a
rational fraction) \cite{Cowan98}. Here we vary the natural frequency
via the current. In Fig.$~$\ref{RFIG8}c we show the
coefficient of variation (CV) of an isolated neuron. When the neuron
is entrained the CV drops to zero. The ISIH then consists
of a single peak. The entrainment occurs for a range of current values,
$I=0.87$-$1.0$. Since the firing rate is constant, the $f-I$ 
has a flat step (not shown). 
Outside the range of entrainment the ISIH
has more structure (Fig.$~$\ref{RFIG8}d and f). 
When there is noise present
in the neuron, the CV will increase. For weak noise
the CV in the entrainment regime will still be lower compared to
the CV when the neuron is not entrained. For the network
the CV is also lower in the entrainment regime (Fig.$~$\ref{RFIG8}b),
though the CV increases
faster with the noise strength compared to the CV of the isolated neuron.
The synchronization of the network, 
measured by $\kappa$ (Fig.$~$\ref{RFIG8}c),
is significantly enhanced in the region of entrainment 
(for $D=0.004$-$0.04$). The enhanced synchronization
disappears for higher values $D>0.2$.

\subsection{Effect of synaptic coupling strength on robustness of strong synchronization}
\begin{figure*}
\newpage
\unitlength=0.1in
\begin{picture}(20,50)
{
\includegraphics{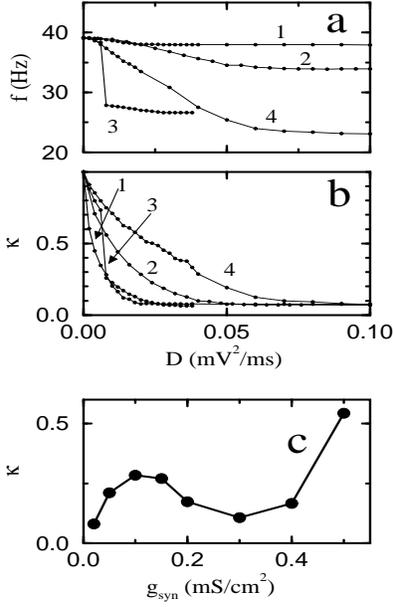}
}
\end{picture}
\caption{(a) Network frequency, $f$, and (b) synchronization parameter, 
$\kappa$, plotted versus noise strength $D$, 
for different values of the synaptic coupling and applied current
($g_{syn}$, $I$)=($0.02$, $0.6955$), ($0.1$, $1.0$),
($0.3$, $1.625$), and ($0.5$, $2.15$),
labeled by (1)-(4), respectively. The value of the
applied current is chosen such that the neuron network
will fire at $f\approx 39.05~Hz$ at $D=0$. In (c) we plot $\kappa$ versus
$g_{syn}$ for $D=0.02$ at the aforementioned current values.
Time averages are computed over $10~s$ after 
a transient of $1~s$,
}
\label{RFIG9}
\end{figure*}

\begin{figure*}
\newpage
\unitlength=0.1in
\begin{picture}(40,50)
{
\includegraphics{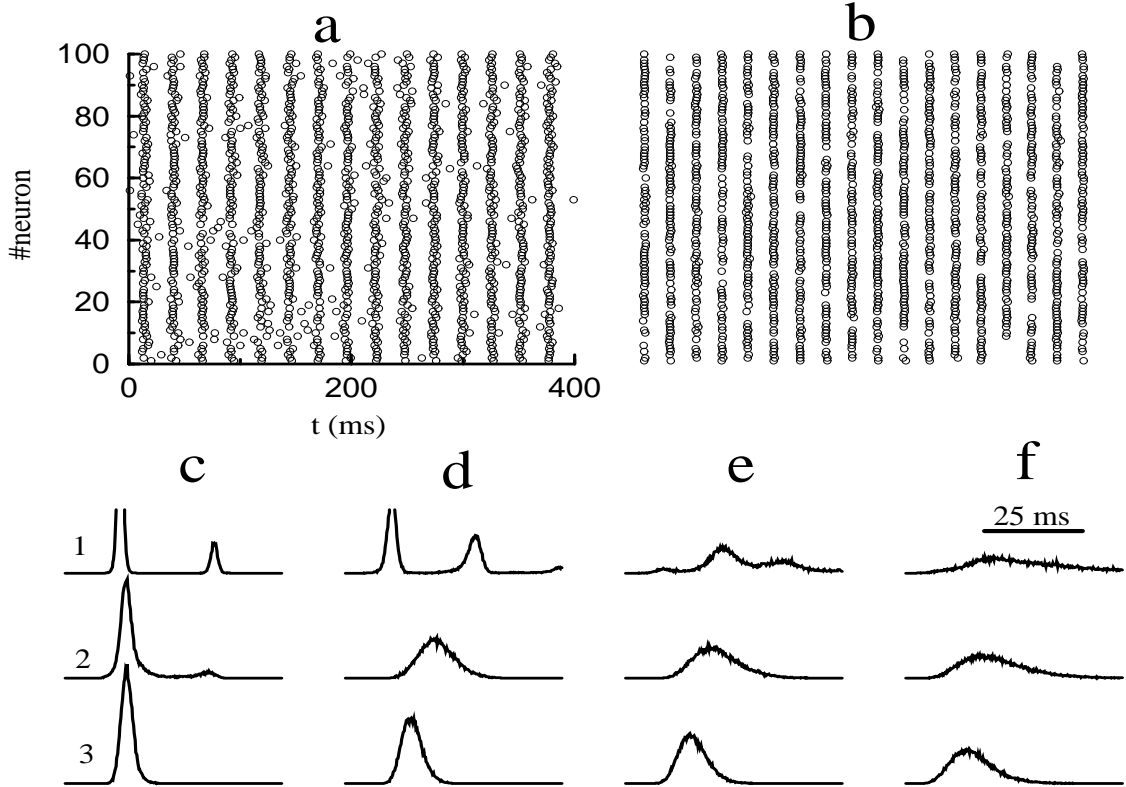}
}
\end{picture}
\caption{
Rastergrams comparison,  (a) $g_{syn}=0.1$, $I=1.0$, and
(b) $g_{syn}=0.5$, $I=2.15$, with noise strength $D=0.002$. 
We also compare the  ISIHs for different values of the
synaptic coupling $g_{syn}=0.5$ (1), $0.2$ (2), and $0.05$ (3),
and for different values of the noise strength $D$ equal to
(c) $0.014$, (d) $0.034$, (e) $0.07$ (e), and (f) $0.17$. The values of the
applied currents are $I=2.15$, $1.331$, and $0.8145$ for 
(1), (2), and (3), respectively. Time-scale bar is shown in f1, the y-scale 
is arbitrary but the same for all curves in (c)-(f). For clarity the top 
of c1 and d1 is cut off. An initial transient of $10~s$
in (a) and (b) was discarded. After a transient of $1~s$, 
time-averages were taken over $10~s$ (c-f). 
}
\label{RFIG10}
\end{figure*}

We have also studied the effect of varying the synaptic coupling strength $g_{syn}$.
For large enough $D$ the network will be asynchronous.
We find that the network frequency in that case decreases with increasing 
values of $g_{syn}$ (Fig.$~$\ref{RFIG9}).
For an asynchronous network the synaptic drive has a constant 
tonic hyperpolarizing conductance, decreasing the firing rate. 
The stronger the  coupling the larger the decrease. The 
synchronization measured by the 
parameter  $\kappa$ displays a different
behavior. In Fig.$~$\ref{RFIG9}c we plot the 
$\kappa$ versus $g_{syn}$ for one specific value $D=0.02$ (and $I$ chosen
such that the firing rate is approximately $39$ Hz). 
It is interesting to note that stronger coupling does not
necessarily mean a higher value of $\kappa$.
The coherence $\kappa$ has a local maximum for $g_{syn}=0.1$, 
for higher values  of $g_{syn}$, $\kappa$ decreases (see WB Fig.$~$12B).
For $g_{syn}>0.3$, $\kappa$ starts increasing again. We have studied
the underlying dynamics of this non monotonous behavior.
In Fig.$~$\ref{RFIG10}c-f we plot the ISIH for different
values of $g_{syn}$. For $g_{syn}\ge 0.2$ one finds more than 
one peak. In the rastergrams (Figs.$~$\ref{RFIG10}a and
(b)), one can see that the dynamics corresponds
to a population that has a well defined frequency,
but individual neurons sometimes miss, or skip, a period. Despite 
this small asynchrony when the neuron fires, it does so
in synchrony with the others.  As a consequence the rastergrams looks 
much more ordered  compared to the one for $g_{syn}=0.1$ at the same
noise strength $D=0.002$. 

\subsection{Larger $g_{syn}$ leads to robust stochastic weak synchronization}
In this subsection we discuss the robust $40$ Hz rhythms found for higher 
$g_{syn}$ values. We have doubled the synaptic decay constant to 
$\tau_{syn}=20$. Here we will use the modified $\kappa_W$ and $CV_W$ 
as mentioned in the Methods section.

\begin{figure*}
\newpage
\unitlength=0.1in
\begin{picture}(20,50)
{
\includegraphics{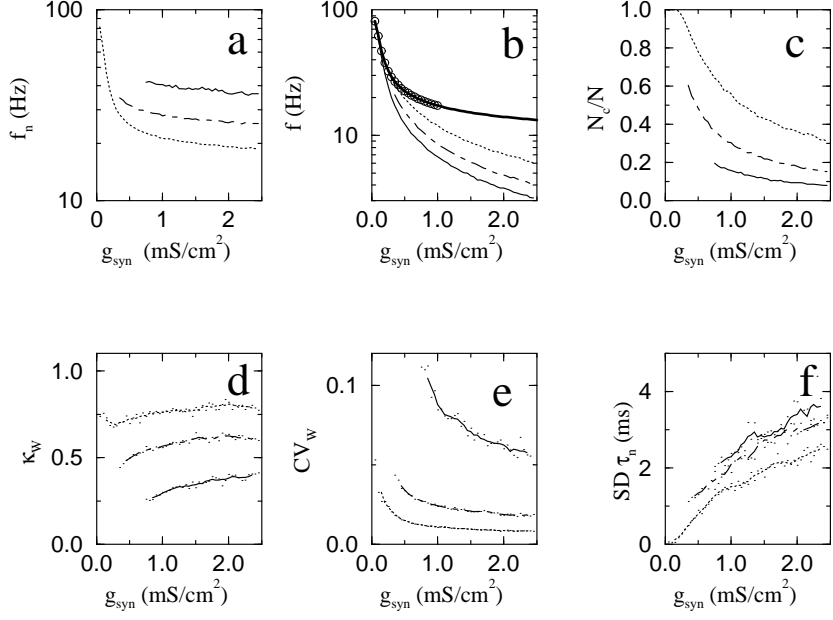}
}
\end{picture}
\caption{ 
We plot (a) network frequency $f_n$, 
(b) single neuron firing rate $f$, (c) cluster size $N_c$ normalized by
system size $N=100$,
(d) $\kappa$ , (e) coefficient of variation CV of ISI, and 
(f) standard deviation of the network period $\tau_n$ versus synaptic strength
$g_{syn}$, for different values of $D=0.0$ (dot-dashed line), $D=0.008$ (dotted)
and $D=0.2$ (solid line). $g_{syn}$ is varied from $0.05$ to $2.5$ in 
increments of $0.05$. For $D=0.008$ ($D=0.2$) the first $6$ ($14$) points 
have been removed
according to the criterium given in the Methods section. In (b) we have added the firing rate of
a single neuron with autosynaptic feedback (fat solid line), and the network
for $D=0$ (open circles).
Other parameters are $\sigma_I=0$, $I=2.0$, $\tau_{syn}=20$. A $500$ ms
transient is discarded, and averages are taken over $2000$ ms. To smooth (d-f)
we have performed running averages over five points, the original points 
are denoted by dots.
}
\label{EXTRA2}
\end{figure*}

In Fig.$~$\ref{EXTRA2} 
we vary $g_{syn}$ from $0.05$ to $2.5$ with a spacing of $0.05$.
The neuron number is kept equal to $N=100$, and we use $I=2.0$,
and $\sigma_I=0$. 
For $D=0.0$ and $\sigma_I=0$ the network is in a strongly synchronized 
state, with the network
frequency $f_n$ the same as the single neuron firing rate $f$. The frequency
is exactly the same as the one for a single neuron with autosynaptic feedback, as one
would expect. This coherent state  can be arrived at from
many different random initial conditions. 

For weak noise, $D=0.008$, 
the network stays in a strongly synchronized state for $g_{syn}<0.25$. 
For higher $g_{syn}$ skipping starts to occur, the fractional cluster size
decreases from values close to one to values below one-half at $g_{syn}=1.2$.
At that point the network is in a real (albeit stochastic) cluster state,
on average the neuron only fires once every two cycles. We will refer to
 all states for which certain active neurons do not fire at each cycle as 
a stochastic weak synchronized (SWS)
network. The network frequency,
$f_n$, and the single neuron firing rate $f$ both decrease with 
increasing $g_{syn}$.
When the network settles in the SWS state $f$ starts to differ considerably
from its value at the $D=0$ state. The strength of synchronization 
increases with $g_{syn}$, that is $CV_W$ decreases and $\kappa_W$ increases. 
For values $g_{syn}>2.0$, $CV_W$ and $\kappa_W$ slowly saturate. 

For stronger noise $D=0.04$ and $D=0.20$ the network is asynchronous
for low values of $g_{syn}$. We have therefore excluded these points based
on the criterium discussed in the Methods section.  The network frequency starts out
at a higher value, and the neurons fire at a lower rate compared
to the $D=0.008$ case. The strength of synchronization, $\kappa_W$ and $CV_W$, 
is reduced compared to the one for $D=0.008$, but still increases with $g_{syn}$. 
Note that all the neurons in the network still have a nonzero firing rate.
It is thus possible to obtain 
weakly synchronized oscillations
in a network consisting of $100$ neurons 
in the frequency range between $20$ and $40$ Hz.

\begin{figure*}
\newpage
\unitlength=0.1in
\begin{picture}(20,50)
{
\includegraphics{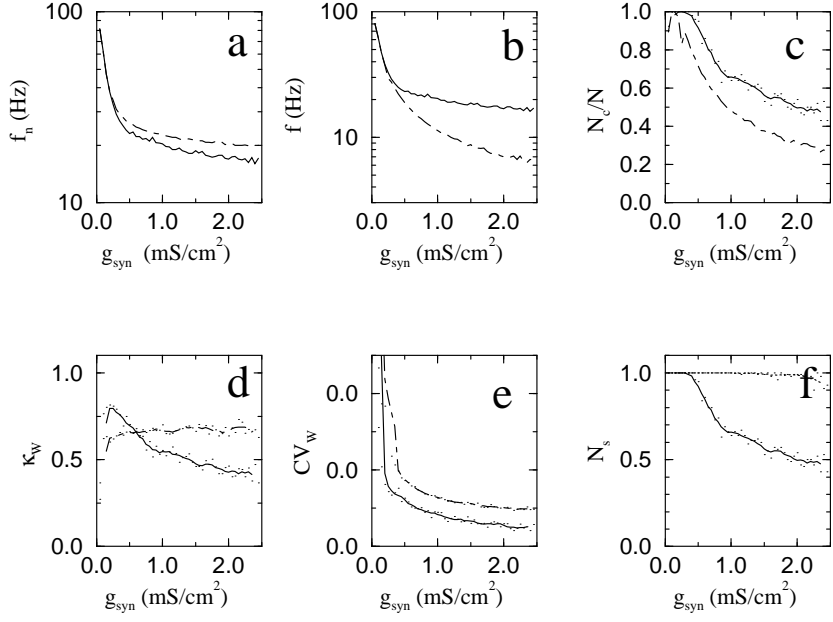}
}
\end{picture}
\caption{ 
We plot (a)-(e) as in the previous figure, 
(f) number $N_s$ of active neurons for different values of 
$D=0.0$ (solid), $D=0.008$ (dot-dashed).
Other parameters are $\sigma_I=0.02$, $I=2.0$, $\tau_{syn}=20$. 
The smoothed (d-f) figures were obtained as in the previous figure.
}
\label{EXTRA3}
\end{figure*}

We find that noise is necessary to obtain SWS.
We have studied SWS in the presence of weak current heterogeneities,
say for $\sigma_I=0.02$. Without noise ($D=0$) the network is in an 
strongly synchronized state, and
$\kappa$ displays a maximum as a function of $g_{syn}$ (WB Fig12B). 
One also clearly notices the effect of suppression \cite{white}:
for larger $g_{syn}$ the inhibition of faster spiking neurons stops
the firing of neurons driven by a smaller current. As a result the total 
number $N_s$
of active neurons gradually drops (Fig.$~$\ref{EXTRA3}f).
For a small amount of noise, $D=0.008$, the situation changes dramatically.
A SWS state is obtained, and all neurons remain active ($N_s/N=1$), while
$\kappa$ saturates for $g_{syn}>0.5$, and the value of $\kappa$ for
 $g_{syn}>0.8$ is even
higher than without noise. Thus noise may actually improve the coincidence.
Of course noise does increase the width, $CV_W$, of the peaks in the 
instantaneous
firing rate. The single neuron firing rate decreases significantly compared
to that for $D=0$, whereas the network frequency is only weakly affected.

We have performed numerical simulations 
for system sizes
$N=10$, $20$, $50$, $100$, $200$, and $1000$. We have used 
the following parameters values: $I=5.0$, $D=0.2$, $g_{syn}=1$, $\tau_{syn}=20$,
and $\sigma_I=0.1$.
The network frequency increases with system size, whereas the firing rate 
stays approximately constant with a dip around $N=50$. The measures
for coherence, $\kappa_W$ and $CV_W$, are also only weakly dependent on
 system size.

\begin{figure*}
\newpage
\unitlength=0.1in
\begin{picture}(20,50)
{
\includegraphics{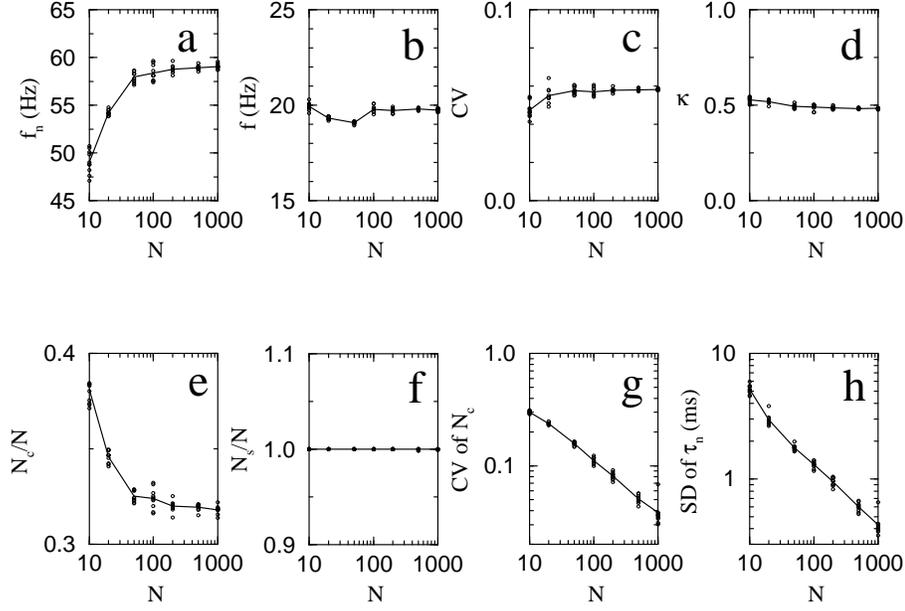}
}
\end{picture}
\caption{ 
We plot (a) network frequency $f_n$, 
(b) single neuron firing rate $f$, 
(c) coefficient of variation CV of ISI,
(d) $\kappa$ , 
(e) cluster size $N_c$ normalized by system size $N$,
(f) number of active neurons $N_s$ normalized by $N$,
(g) coefficient of variation of $N_c$,
(h) standard deviation of the network period $\tau_n$ 
versus system size $N$ ($N=10$, $20$, $50$, $100$, $200$, $500$, $1000$).
We show results for ten different realizations of the current
distribution (open circles), and the their average (solid lines).
Other parameters are $\sigma_I=0.10$, $D=0.2$, $I=5.0$, $g_{syn}=1.0$, 
and $\tau_{syn}=20$. A $500$ ms
transient is discarded, and averages are over at least $2000$ ms. 
}
\label{EXTRA4}
\end{figure*}

The cycle-to-cycle fluctuations in cluster size vary approximately
as $\sqrt{N}$ (Fig.$~$\ref{EXTRA4}g). 
The strength of the inhibition is determined by the
number of neurons that fired in the previous cluster, and
in turn it determines at what time the first neurons
become disinhibited. One therefore expects cluster size fluctuations
and cycle length fluctuations to be intimately related.
Indeed, the standard deviation of cycle length varies as $1/\sqrt{N}$ with
$N$ the number of neurons (Fig.$~$\ref{EXTRA4}h).
This means that  larger networks are better at generating a precise cycle 
length,
whereas size does not matter as much for the coincidence of spikes
measured  by $\kappa_W$ and $CV_W$.

In each simulation we randomly draw a set of driving currents $I_j$ for
each neuron $j$ from a uniform probability distribution. The results one obtains may 
critically depend on the
particular realization of driving currents. One expects that for 
larger systems this is less of a problem. The population distribution
of $I$ is more likely to approach the original ensemble distribution 
of currents for a given neuron.
Here we have studied the range of values for the measured quantities 
($f_n$, $f$, and so on) 
for ten different realizations. We find that for most 
quantities (for these
parameter values) the range of values decreases with $N$, 
and for $N\ge 500$ one realization will give a result close 
to the expectation value.

\begin{figure*}
\newpage
\unitlength=0.1in
\begin{picture}(20,50)
{
\includegraphics{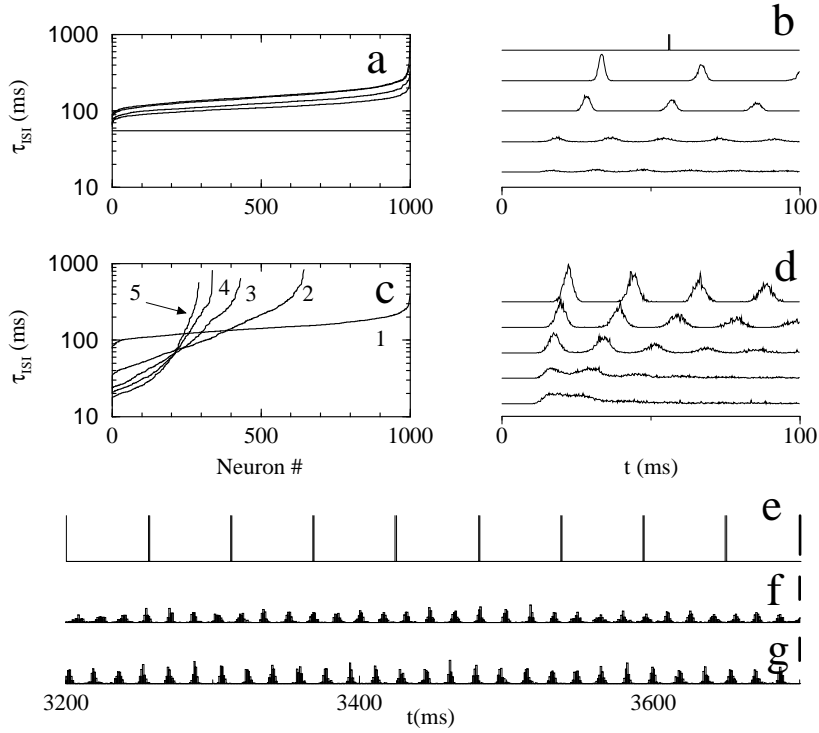}
}
\end{picture}
\caption{We plot in (a,c) the interspike intervals (ISI) of the network neurons 
sorted from the lowest to highest value, and (b,d) the ISIHs.
In (a,c) we used different values of the noise-strength, from top to bottom
$D=0.0$, 
$0.04$, $0.08$, $0.36$, and $0.56$, with $\sigma_I=0$. In (c,d) 
from top to bottom $\sigma_I=0$ (1), $0.173$ (2),$0.346$ (3), $0.520$ (4), 
and $0.693$ (5),  with $D=0.2$.  
After a transient of $500$ ms, the time-average is computed over $2~s$.
We plot the instantaneous
firing rate as a function of time for (e) $D=0$, $\sigma_I=0$,
(f) $D=0.56$, $\sigma_I=0$, and (g) $D=0.2$, $\sigma_{I}=0.35$. The scale
bars are (g) $500$ imp/s, (h,i) $50$ imp/s. 
We used $I=3.5$, $\tau_{syn}=20$, 
$N=1000$, and $g_{syn}=2.0$.
}
\label{EXTRA5}
\end{figure*}

\begin{figure*}
\newpage
\unitlength=0.1in
\begin{picture}(20,50)
{
\includegraphics{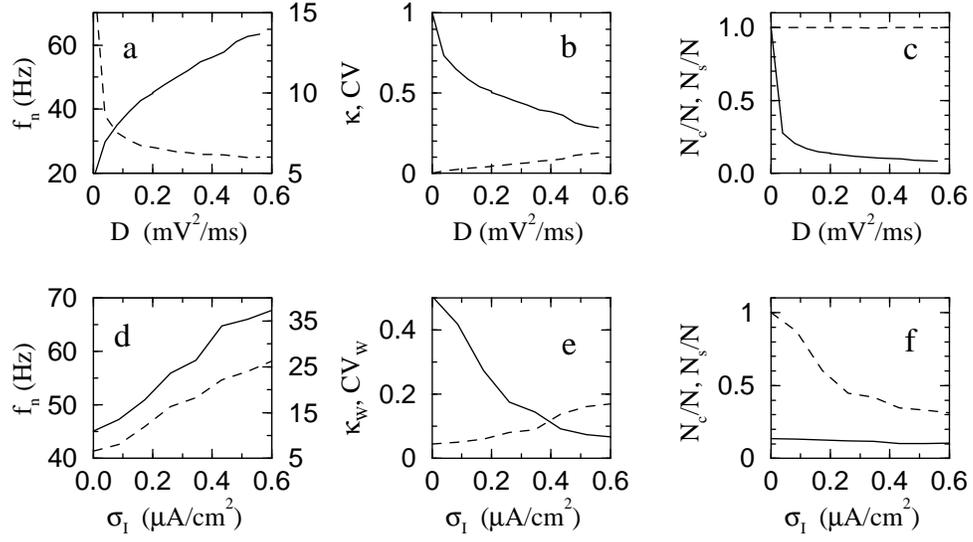}
}
\end{picture}
\caption{ 
We plot in (a,d) the network frequency $f_n$ (solid line, left hand scale)
and single neuron firing rate $f$ (dashed line, right hand scale);
(b,e) coherence $\kappa$ (solid line) and the CV (dashed line);
(c,f) cluster size $N_c$ and number of active neurons $N_s$ divided by $N$; 
as function of (a,b,c) noise variance $D$,
and (d,e,f) current heterogeneity $\sigma_I$. 
We used the  following
parameters $I=3.5$, $\tau_{syn}=20$, $g_{syn}=0$,  and $N=1000$,
for (a,b,c) with $\sigma_I=0$
 and for (d,e,f) the same parameters with $D=0.2$. 
}
\label{RFIG15}
\end{figure*}

We now vary $\sigma_I$ and $D$ for the following fixed parameter set
$N=1000$, $g_{syn}=2$, $\tau_{syn}=20$, and $I=3.5$. 
For $D=0$ and $\sigma_I=0$ the network is strongly synchronized
at $20$ Hz. The  instantaneous firing rate consists of a sequence
of regularly spaced delta functions (Fig.$~$\ref{EXTRA5}e),
the ISIH has a single delta peak at $50$ ms (Fig.$~$\ref{EXTRA5}b), and
all neurons spike at the same frequency (Fig.$~$\ref{EXTRA5}a).
Increasing $D$ increases the network frequency, but decreases the
single neuron firing rate (Fig.$~$\ref{RFIG15}). The population activity is
 still periodic (Fig.$~$\ref{EXTRA5}f), but the peaks have a finite width
(as well as the ISIH), and the ISIH becomes multimodal. This process
continues with ISIH spreading out more and more, with the $CV_W$ increasing,
and $\kappa_W$ decreasing.

As mentioned before we need some noise to generate  an SWS state.
Here we use $D=0.2$, while at the same time varying $\sigma_I$. For 
finite $\sigma_I$
there is still a coherent population activity (Fig.$~$\ref{EXTRA5}g),
despite the fact that neurons have different firing rates (Fig.$~$\ref{EXTRA5}c).
Increasing $\sigma_I$ will reduce coherence, $\kappa_W$ decreases and
$CV_W$ increases. At the same time both $f_n$ and $f$ increase 
(Fig.$~$\ref{RFIG15}d).
This is different from the effect of increasing $D$. Higher 
$\sigma_I$ leads to 
suppression, with fast spiking neurons preventing slower ones from firing, and as
a result part of the inhibition disappears, while further increasing 
the firing rate
and its average (calculated from the active neurons).
On the other hand noise increases the tonic inhibition for each 
neuron, and thus leads to
a reduced firing rate. Also the progression of the asynchronous state 
is different.
The first peak in the ISIH becomes broader, and the higher order
ones have a reduced prominence (Fig.$~$\ref{EXTRA5}d). For increasing $D$ the
peaks just wash out.

\subsection{Comparison to externally induced stochastic weak synchronization} 
\begin{figure*}
\newpage
\unitlength=0.1in
\begin{picture}(40,50)
{
\includegraphics{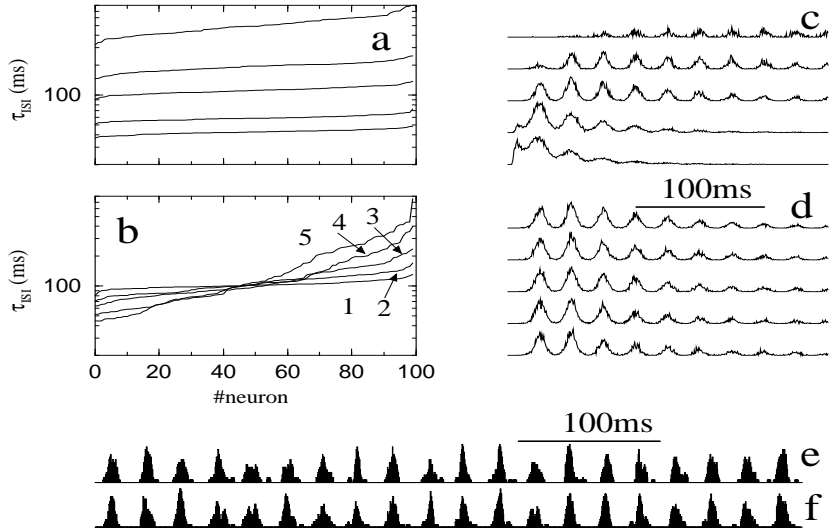}
}
\end{picture}
\caption{We plot in (a,b) the interspike intervals (ISI) for the $100$ neurons 
sorted from the lowest to highest values, and in (c,d) the ISIHs.
In (a,c) we used different values for the noise-strength, from top to bottom
 $D=0.04$, $0.24$, $0.84$, $4.2$, and $8.2$, with $\sigma_I=0$. In (b,d) 
from top to bottom
$\sigma_I=0$ (1), $0.04$ (2),$0.08$ (3), $0.12$ (4), and $0.16$ (5), with 
$D=1.04$. A sinusoidal current with a period of $25~\mbox{ms}$ and an 
amplitude $1.2~\mu A/cm^2$ is added to the soma. The time-scale
bar is shown in d1, the y-scale is arbitrary but the same in (c), and (d). 
After a transient of $2~s$, the time-average is computed over $10~s$.
(e,f) Instantaneous spiking rate versus time for (e) $\sigma_I=0$,
and (f) $\sigma_I=0.10$, both at  $D=1.04$.
}
\label{RFIG12}
\label{RFIG13}
\end{figure*}

We find that 
robustness can be enhanced
by getting a network in a SWS state.
The single neuron discharge in an SWS state is very similar 
to that obtained in
Stochastic Resonance \cite{Wies95} using a subthreshold {\it external} drive. 
In this subsection we compare the previous self-induced SWS state
to the one induced by an external drive. 
We drive the system by a sinusoidal current of amplitude $1.2$ (admittedly
large). The average value of the driving current is zero.
For weak noise the spikes of a single neuron are spaced many
cycles apart (Fig.$~$\ref{RFIG12}a).  With increasing noise
the ISIH starts to look more like the ones in Fig.$~$\ref{EXTRA5}b. The
 population activity
also is periodic (Fig.$~$\ref{RFIG12}e). The distribution
of $\tau_{ISI}$ in Fig.$~$\ref{RFIG12}a looks similar to the one
in Fig.$~$\ref{EXTRA5}a. For high values of $D$ the neuron
can spike more than once during one cycle, as a result the
number of coincident spikes is reduced.
The addition of current heterogeneity does not seem to affect
the network behavior (as long as 
all the neurons are still below threshold). The different ISIH
in Fig.$~$\ref{RFIG12}d look very similar. However the corresponding 
range of $\tau_{ISI}$
values in Fig.$~$\ref{EXTRA5}c does increase with $\sigma_I$.

\section{Discussion}
Previous authors have recognized that strong synchronization
is only moderately robust against neuronal heterogeneity \cite{wang,white}.
We have previously shown that the same holds including synaptic noise \cite{CNS}.
The basic premise of synchronization by mutual inhibition is
almost trivial, since the network consists of intrinsically periodically
spiking neurons. Their output produces a periodic synaptic drive,
which in turn is fed back into the network. Inhibition thus allows
a phase lock at zero phase with this drive. Heterogeneity
and noise reduces the phasic, and increases the tonic part
of the synaptic drive, leading to a reduction in synchronization, and
eventually leading to an asynchronous state (Fig.$~$\ref{RFIG5}c,d).
The synchronization
behavior of networks of physiological realistic neurons, however, is by no means fully 
understood.
In this work we showed that the loss of synchronization
proceeds via {\bf different} mechanisms in the presence of synaptic noise
compared to the presence of current heterogeneity.
This is evident from the cross correlations shown
in Figs.$~$\ref{RFIG6} and $~$\ref{EXTRA1}.
We also found that the noise-induced precision loss
in the uncoupled neuron is exacerbated by the inhibitory coupling.
All of these could seem obvious based on previous  
work on heterogeneity \cite{wang,white}. However, its consequences for real
life biological networks had not been fully appreciated.
Our results, combined with previous results, show that there
is a problem with strong synchronization by mutual inhibition, since
it is unlikely to occur in {\it in vitro} or {\it in vivo} systems.
[There are exceptions such as for example the pacemaker nucleus
in electric fish \cite{Moortgat98}, where the neurons are coupled via 
gap junctions.]
The aim of this paper was to treat this problem: how can
one obtain robust synchronization in the presence of
synaptic noise and neuronal heterogeneity?
Our results are twofold. First, methods to increase
the robustness of strong synchronization have been ineffective.
Second, we showed that robust stochastic weak synchronization (SWS) can be obtained
for biophysically realistic parameter values. SWS is consistent with
previous experimental data.
In what follows we discuss these two important results
in more detail. 

We believe that strong synchronization
is not robust enough. To make sure that we do 
not prematurely discard strong synchronization
by mutual inhibition we have to make an effort to
increase robustness.
In this paper we discussed two simple methods
to increase robustness of strong synchronization.
One method was to increase the synaptic coupling $g_{syn}$.
Inhibition is responsible for synchronization. 
It is then quite natural to expect that increasing the strength
of inhibition increases robustness. The fact that this does not
happen is surprising. For current heterogeneity this
is in part due to suppression \cite{white,wang}. 
We have studied this effect for synaptic noise in more detail.
We found that neurons skip periods for higher values of $g_{syn}$ 
(see Fig.$~$\ref{RFIG10}).
In other words the strongly synchronized state becomes unstable,
and a weakly synchronized state emerges. This
weakly synchronized state looked more coherent (Fig.$~$\ref{RFIG10}a,b), and
provided the impetus to further study the robustness of the SWS states.

We also added a periodic drive to the neuron.
Recent experimental work shows that the CV of neurons
on an entrainment step is reduced compared to the CV outside
the step \cite{Cowan98}. 
A clear physiological correlate in hippocampus of this drive is lacking 
at present.
Since we try to reject our conjecture this lack of physiological realism
is not a problem. We found a moderate increase in robustness 
(see Fig.$~$\ref{RFIG8}).
The periodic drive is not as effective as one would have intuited, however. 
In fact the inhibitory connections reduce the increase in robustness compared
to the increase in the single uncoupled neuron (see Fig.$~$\ref{RFIG8}).
If we add a subthreshold periodic drive with
noise to a quiescent neuron we obtain weak synchronization. This is 
known as Stochastic Resonance in excitable systems \cite{Wies95}.
To summarize:  our attempts to significantly increase robustness of strong 
synchronization failed. Instead we found weak 
synchronization. Thus it is easier to find weak synchronization in parameter
space than it is to find strong synchronization. 

If one accepts the fact, however, that strong synchronization
is not robust against noise and heterogeneity, and that periodic
population oscillations are found in experiments, then one has to
carefully consider the possible relevance of weak synchronization.
Weak synchronization as well as strong synchronization lead to a periodic 
population discharge,
and specifically to an inhibitory synaptic drive indistinguishable from the one found
in pyramidal neurons in \cite{whittington}.
Moreover, the clusters that form in stochastic weak synchronization
bear a resemblance to the neuronal assemblies found
in some experiments \cite{Laurent96a}, and that are thought to play
a role in putative binding \cite{binding3}.
The question then is, is SWS more robust, and can it be found
for the gamma frequency range for biophysically realistic parameters?
What is needed is a higher
total synaptic conductance, and noise. The necessary amount of noise
is very small, $D>0.004$ is sufficient. The noise prevents the occurrence
of suppression (Fig.$~$\ref{EXTRA3}). In suppression the faster neurons
prevent the slower ones from firing. This reduces the inhibition of the 
faster ones, and allows them to fire at different frequencies and at 
random relative phases. Suppression is thus detrimental to synchronization.

We obtained SWS for different system sizes (we studied
networks from $10$ to $1000$ neurons). The coincidence properties
($\kappa_W$, and $CV_W$) did not vary much with size. The
temporal precision of the population oscillation, however, increases
approximately as $1/\sqrt{N}$ (Fig.$~$\ref{EXTRA4}h). Large networks can thus
 produce
precise pacemaker rhythms.  In addition the statistical quantities
in small networks show more variation with different realizations of
the current drive. 
It is of considerable interest to understand why weak synchronization
is so much more robust and prevalent compared to strong synchronization.
In strong synchronization one requires an equal firing rate for each neuron,
and coincident spikes, while weak synchronization requires only coincident
spikes. By definition, then, weak synchronization is easier to generate.
In fact strong synchronization is intrinsically less robust, for it requires
phase locking between neuron pairs. This is only possible
(depending on intrinsic properties) for a small difference in driving
currents. There is a price to pay, there will be a small phase difference 
between the firings of each neuron. Pairs with a large phase difference are
less stable against the influence of noise. It is instructive to consider this
problem as the nonlinear dynamics of a single neuron driven by 
a periodic drive. 
The neuron can be  entrained on different n:m steps.
On these steps in the $f$-$I$ plot the neuron generates $n$
 action potentials during
$m$ cycles of the periodic drive for a range of values of $I$ 
(or intrinsic frequency). 
Noise induces jitter in the spike time, but the CV on the steps is reduced
compared to the CV outside the steps (and steps with higher $m$ values). 
The entrainment produces a phase difference between the firing time
and the crest of the drive. The size of the phase difference depends on the
intrinsic frequency of the neuron. Robustness is less for current values
close to the edge of the step (with unfavorable phase differences).
Therefore noise will reduce the width of the phase-locking step (not shown). 
Here we found network parameters for which coincidence could be maintained 
despite highly variable firing rates of the neuronal populations.

Our work, and also a recent study \cite{white},
is to a large extent based on the recent contributions by Wang\&Buzsaki 
\cite{wang}.
It is therefore important to briefly reiterate, and spell out how our work 
extends the work of Wang \& Buzsaki,
and how it differs \cite{white}.
We included the effect of synaptic noise. We have shown that for the
purposes of our modeling work a Gaussian white noise current
can adequately reproduce experimental ranges of CV \cite{CNS98}.
Biophysically realistic amounts of noise do affect the synchronization
we have studied. The noise effects are also different from those of current
heterogeneity as was discussed before. The issue of noise was not addressed
by Wang \& Buzsaki.
Another important difference is that previous works \cite{wang,white}
studied only strong synchronization. Here we have proposed that stochastic
weak synchronization underlies the synchronized population oscillations
in the hippocampus. For this reason our
robust 40Hz population rhythms were obtained for
different values of the coupling parameters $g_{syn}$, $\tau_{syn}$, 
and the driving current $I$, and the system size $N$
compared to previous work\cite{wang}. 
In our computational work we actually needed a small amount of noise to obtain 
weak synchronization.

The synchronization properties of
large networks
may be of some mathematical interest. Our networks are small, 
and probably the behavior can change
quantitatively when increasing the  network size significantly.
However, in this paper we only addressed the 
question as to whether networks of physiologically realistic size and connectivity 
can robustly synchronize. Recent experimental work suggests that 
interneurons contact
on the order of $60$ other interneurons \cite{Sik95}. For that reason we
only vary our network size between $N=10$ and $1000$.

In the introduction we mentioned recent {\it in vivo} work
and the controversies on the functional role of synchronization.
Our work obviously does not contribute to the understanding of the function
of synchronization. An important question is what kind of synchronization
can be sustained in biophysically realistic networks.
Traub and coworkers \cite{traub,traub2,whittington,Traub98a} looked for
physiological correlates of the gamma rhythms using {\it in vitro}
and computational experiments. Their results show the crucial role of 
inhibition, 
and have provided much of the impetus for our work. 
The nervous system produces, for some unknown reason, periodic population
activity using circuitry consisting of noisy and heterogeneous neurons.
Our results establish that it
is possible for inhibitory neurons to be the driving
force for synchronization under these conditions.

\section{Acknowledgments}
This work was partially funded by the Northeastern University CIRCS fund,
and the Sloan Center for Theoretical Neurobiology (PT).
We thank W-J Rappel for help during the initial stage of this
work, and TJ Sejnowski for useful suggestions. 
Part of the calculations were performed at Northeastern University
High Performance Computer Center.


\onecolumn
\newpage

\newpage

\newpage

\newpage

\newpage

\newpage
\newpage

\newpage

\newpage

\newpage

\newpage
\newpage

\end{document}